\documentclass[10pt]{article}
\def\D{\Delta}
\def\L{\Lambda}
\def\l{\lambda}
\def\S{\Sigma}
\def\g{\gamma}
\def\e{\epsilon}
\def\s{\sigma}
\def\dim{\textrm{dim}}
\newcommand{\be}{\begin{equation}}
\newcommand{\ee}{\end{equation}}
\newcommand{\bea}{\begin{eqnarray}}
\newcommand{\eea}{\end{eqnarray}}

\begin{document}

\begin{center}
\bf{SPIN FOAM MODELS OF QUANTUM GRAVITY}\footnote{Talk at the
Balkan Workshop 2003, 29 Aug. - 2. Sept., Vrnja\v cka Banja,
Serbia. Work supported by the FCT grants POCTI/FNU/49543/2002 and
POCTI/MAT/45306/2002.}
\end{center}

\bigskip
\bigskip
\begin{center}
A. MIKOVI\'C
\end{center}

\begin{center}\textit{Departamento de Matem\'atica e Ci\^encias de Computac\~ao \\
Universidade Lus\'ofona de Humanidades e Tecnologias\\
Av. do Campo Grande, 376, 1749-024 Lisbon, Portugal\\
E-mail: amikovic@ulusofona.pt}
\end{center}

\bigskip
\bigskip
\begin{quotation}
\small{We give a short review of the spin foam models of quantum
gravity, with an emphasis on the Barret-Crane model. After
explaining the shortcomings of the Barret-Crane model, we briefly
discuss two new approaches, one based on the 3d spin foam state
sum invariants for the embedded spin networks, and the other based
on representing the string scattering amplitudes as 2d spin foam
state sum invariants.}\end{quotation}

\bigskip
\bigskip
\noindent{\bf{1. Introduction}}

\bigskip
\noindent The spin foam models originate from the Ponzano-Regge
model of 3d Euclidian quantum
gravity \cite{pr}. The idea there was to use the simplical complex, i.e.
the spacetime triangulation,
whose triangles had integer lengths, which were proportional to the
spins of the $SU(2)$ group. Then the 3d gravity path integral (PI) was defined
as a sum over the spins of the products of the $6j$ symbols which were
associated to the tetrahedrons of the simplical complex. A length cut-off
was introduced in order
to regularize the path integral. Since 3d gravity
is a topological theory, the corresponding path integral would be
a topological invariant of the 3d manifold. However, the
Ponzano-Regge path integral was not a topological invariant,
because the topological invariance required the quantum $SU(2)$
group at a root of unity, which was discovered by
Turaeev and Viro \cite{tv}.

Still, the idea of the Ponzano-Regge model was useful, because
the model can be understood as a path integral for the $SU(2)$ BF
theory \cite{x}. This then inspired Ooguri to consider a 4d
version of the model, as a PI for the 4d $SU(2)$ BF theory
\cite{o}. In this case the areas of the triangles are integer
valued, i.e. proportional to the spins. The PI was
formally topologically invariant, but divergent. A well-defined
topological invariant was obtained by Crane, Yetter and Kauffman
\cite{cyk}, who replaced the $SU(2)$ group by the quantum $SU(2)$ group
at a root of unity, so that the $SU(2)$ spins become bounded by a maximal spin.
However, the
corresponding invariant was not new, contrary to the 3d case, since
it gave a signature of the 4-manifold.

At the same time, Baez proposed the idea of the spin foams
\cite{b}, as a way of understanding the results of loop quantum
gravity \cite{lqg}, from a spacetime perspective, i.e. a spin foam
is a time evolution surface of a spin network, so that the spins
are naturally associated to the faces of the spin foam. Given that
the Einstein-Hilbert (EH) action of GR can be understood as a
constrained BF theory, this then prompted Crane and Barrett to
look for a constrained version of the CYK topological spin foam
state sum \cite{ebc,bc}.

\bigskip
\bigskip
\noindent{\bf{2. The Barret-Crane model}}

\bigskip
\noindent The EH action can be written as the $SO(3,1)$ BF theory
action \be \int_M Tr (B\wedge F) \quad,\ee where
$F_{ab}=d\o_{ab} + \o_a^c\wedge\o_{ca}$
is the curvature two-form for the spin connection $\o$, and the two-form
$B$-field is constrained by
\be B_{ab}=\epsilon_{abcd}e^c \wedge e^d \quad,\label{bcon}\ee
where the $e$'s are the tetrad one-forms. The BF
theory path integral can be written as \bea Z &=& \int DA\,DB\,
\exp\left(i\int_M Tr(B\wedge F)\right)\nonumber\\ &=& \int \prod_l
dA_l \prod_\D dB_\D \exp\left(i\sum_{f} Tr(B_\D F_f
)\right)\quad,\eea where $l$ and $f$ are the edges and the faces
of the dual two-complex $\cal F$ for the simplical complex $T(M)$, while
$\D$ are the triangles of $T$. The variables $A_l$ and $B_\D$ are defined
as $\int_l A $ and $\int_\D B $ respectively, while $F_f = \int_f F$.

By performing the $B$ integrations
one obtains \be Z= \int \prod_l dA_l \prod_f \delta (F_f) \quad,\ee
which can be defined as \be Z= \int \prod_l dg_l \prod_f \delta
(g_f) \quad,\ee where $g_f = \prod_{l\in\partial f} g_l$. By using
the well-known identity
\be \delta(g) = \sum_\L \textrm{dim}\,\L \,\chi_\L (g) \quad,\ee
where $\L$'s are the irreducible representations (irreps) of the group and
$\chi$'s are the characters,
one obtains
\be Z= \sum_{\L_f,\iota_l} \prod_f \dim\,\L_f \prod_v A_v
(\L_f,\iota_l) \quad,\label{tsfss}\ee
where $A_v$ is the vertex amplitide associated to the 4-simplex dual to
the vertex $v$. This amplitude is given
by the evaluation of the corresponding 4-simplex spin network, known as the
$15j$ symbol.
The sum (\ref{tsfss}) is called a spin foam state sum, because it is a sum of
the amplitudes for the colored two complex $\cal F$, i.e. a spin foam.

One can now conjecture that exists a quantization procedure such
that the quantities $B_\D$ become the 4d rotations algebra
operators $J_\D$, since the 4d rotation group irreps are labelling
the triangles $\D$, or the dual faces $f$. Then one can show that
the constraint (\ref{bcon}) becomes a constraint on the
representations labelling the triangles $\D$, given by
\be\e^{abcd}J_{ab}J_{cd}=0 \label{simp}\ee \cite{ebc,bc}. In the
Euclidian case the irreps are given by the pairs of the $SU(2)$
spins $(j,j')$, so that the constraint (\ref{simp}) implies
$j=j'$. In the Minkowski case, requiring the hermiticity of the
$B$ operators implies that one needs the unitary irreps of the
Lorentz group. These are infinite-dimensional irreps and they are
given by the pairs $(j,p)$ where $j$ is the $SU(2)$ spin and $p$
is a continuous label. The constraint (\ref{simp}) implies that
$\L = (0,p)$ or $\L=(j,0)$.

One can argue that the spacelike triangles should be labelled by
the $(0,p)$ irreps, while the time-like triangles should be
labelled by the $(j,0)$ irreps. Since a spacetime triangulation
can be built from the spacelike triangles, Barrett and Crane have
proposed the following spin foam state sum (integral) for the
quantum general relativity \cite{bc}\be Z_{BC} = \int \prod_f p_f
dp_f \prod_v \tilde A_v (p_f) \quad, \label{bcss}\ee where $\tilde
A_v$ is an amplitude for the corresponding 4-simplex spin network,
given by \be \tilde A (p_1,\cdots,p_{10})= \int_{H^5}
\prod_{i=1}^5 dx_i \delta(x_1 -x_0) \prod_{i<j}K
_{p_{ij}}(x_i,x_j)\quad.\ee This is as an integral over the fifth
power of the hyperboloid $H=SO(3,1)/SO(3)$ of a propagator $K_p
(x,y)$ on that space. The propagator is given by \be  K_p (x,y)=
{\sin\left( p d(x,y)\right)\over p \sinh d(x,y)}\quad, \quad \cosh
d(x,y)= x\cdot y \quad.\ee

The expression (\ref{bcss}) is not finite for all triangulations,
but after a slight modification, consisting of including a
non-trivial edge amplitude $\tilde A (p_1,\cdots,p_4)$, the
partition function becomes finite for all non-degenerate
triangulations \cite{cpr}. This was a remarkable result, because
it gave a perturbatively finite\footnote{$Z_{BC}$ depends on a
triangulation, in accordance with the fact that 4d gravity is
non-topological, and hence one should also sum over the
triangulations in order to obtain a well-defined quantity. How to
do this it is not clear at present, so that one can obtain only
the perturbative results.} quantum theory of gravity, which was
not based on string theory.

The main difficulties with the BC type models are:

1) It is difficult to see what is the semi-classical limit, i.e.
what is the corresponding effective action, and is it given by the EH
action plus the $O(l_P)$ corrections, where $l_P$ is the Planck length.

2) Coupling of matter: since matter couples to the gravitational
field through the tetrads, one would need a formulation where a
basic field is a tetrad and not the $B$ 2-form. In the case of the
YM field, the coupling can be expressed in terms of the $B$ field
\cite{amm}, so that one can formulate a BC type models
\cite{amym,y}. However, for the fermions this is not possible, and
a tetrade based formulation is necessary. In \cite{amm} an
algebraic approach was proposed in order to avoid this problem,
and the idea was to use a result from the loop quantum gravity,
according to which the fermions appear as free ends of the spin
networks. Hence including open spin networks gives a new type of
spin foams \cite{amnsf}, and this opens a possibility of including
matter in the spin foam formalism. However, what is the precise
form of the matter spin foam amplitudes remains an open question.

\bigskip
\bigskip
\noindent{\bf{3. New directions}}

\bigskip
\noindent Given the difficulties of the BC model, we have proposed
two new directions how to use the spin foam state sum formalism in
order to arrive at a desirable quantum theory of gravity.

In \cite{amlqg} it was proposed to use the 3d spin foam state sum
invariants in order to define the relevant quantities in the loop
quantum gravity formalism. The idea is to use the representation
of a quantum gravity state $|\Psi\rangle$ in the spin network
basis \be |\Psi\rangle = \sum_\g |\g\rangle\langle\g |\Psi\rangle
\quad.\ee The expansion coefficients are then invariants of the
embedded spin networks in the spatial manifold $\S$, and can be
formally expressed as \be \langle\g |\Psi\rangle = \int DA\,
\langle\g |A\rangle\langle A|\Psi\rangle = \int DA\, W_\g
[A]\,\Psi[A] \quad,\ee where $A$ is a 3d complex $SU(2)$
connection, $W_\g [A]$ is the spin network wave-functional
(generalization of the Wilson loop functional) and $\Psi[A]$ is a
holomorphic wave-functional satisfying the quantum gravity
constraints in the Ashtekar representation.

In the case of non-zero cosmological constant $\l$, a non-trivial
solution is known, i.e. the Kodama wavefunction \be \Psi[A] =
e^{\frac{1}{\l}\int_\S Tr\left(A\wedge dA +\frac23 A\wedge A\wedge
A\right)} \quad,\ee while in the $\l=0$ case a class of formal
solutions is given by \be \Psi[A] = \prod_{x\in\S}\delta
(F_x)\,\Psi_0 [A] \quad,\ee i.e. a flat-connection wavefunction
\cite{amlqg}. In the $\l=0$ case one can show that the
corresponding spin network invariant is given by a 3d spin foam
state sum for the quantum $SU(2)$ at a root of unity \cite{amlqg}.

In the $\l\ne 0$ case, it is conjectured that the corresponding
spin network invariant is given in the Euclidian gravity case by
the Witten-Reshetikhi-Turaeev invariant for $q=e^{2\pi i/(k+2)}$,
where $k\in\bf N$ and $\l =k/l_P^2$, while in the Minkowski case,
the invariant is given by an analytical continuation of the
Euclidian one, as $k\to ik$ \cite{smo}.

In \cite{amst} it was proposed to use the 2d spin foam state sums
in order to define a string theory as a quantum theory of gravity.
The main idea is to use the string theory formal expression for
the scattering amplitude of $n$ gravitons (or any other massless
string modes), given as \be A (p_1,...,p_n) = \int_\S d\s_1 \cdots
\int_ \S d\s_n \,\langle V_{p_1}(\s_1)\cdots V_{p_n}(\s_n )\rangle
\quad,\label{ssa}\ee where \be \langle V_{p_1}(\s_1)\cdots
V_{p_n}(\s_n )\rangle =\sum_{2d\, metrics}\int DX\,e^{i\int_\S
(\partial X )^2}\, V_{p_1}(\s_1)\cdots V_{p_n}(\s_n )\quad,\ee and
argue that (\ref{ssa}) should represent a 2d BF theory ivariant
for the $\theta_n$ spin network embedded in the string world-sheet
manifold $\S$. The BF theory group is given by the group of
isometries of the spacetime background metric, and in \cite{amst}
a simple possibility for the amplitude was considered \be
A(p_1,...,p_n)=\int DA\, DB\, e^{i\int_\S Tr(B\wedge
F)}\,W_{\theta_n}[A] \label{tst}\quad,\ee where the isometry group
was taken to be $SU(2)$. Then the labels $p_1,...,p_n$ become the
$SU(2)$ spins, and there is a maximal spin, because the PI
(\ref{tst}) becomes a state sum for the quantum $SU(2)$ at a root
of unity. Because the $BF$ theory is a topological theory, one can
expect that the amplitude (\ref{tst}) will correspond to a
topological string theory.

\end{document}